\documentclass[12pt,a4]{article}
\usepackage{a4wide}
\usepackage{amsfonts}
\usepackage{amstext}
\usepackage{amsmath}
\usepackage{amssymb}
\usepackage{amsthm}
\usepackage[mathscr]{eucal}
\usepackage{graphicx}
\usepackage{amsmath}
\usepackage{hhline}

\def\llangle{\langle}
\def\rrangle{\rangle}

\def\C{\mathsf{C}}
\def\D{\mathsf{D}}
\def\E{\mathsf{E}}
\def\DD{D}
\def\EE{E}
\def\dd{\mathsf{d}}
\def\ee{\mathsf{e}}
\def\1{\mathsf{1}}
\def\({\left(}
\def\){\right)}
\def\Z{\mathbb{Z}}
\def\Q{q^{-1}}
\def\inv{^{-1}}

\begin{document}

\title{
Asymmetric Simple Exclusion Process with Open Boundaries and 
Askey-Wilson Polynomials\\
}

\author{
Masaru UCHIYAMA
{\footnote {\tt E-mail: uchiyama@monet.phys.s.u-tokyo.ac.jp }}\, ,
\setcounter{footnote}{2}
Tomohiro SASAMOTO
{\footnote {\tt E-mail: sasamoto@stat.phys.titech.ac.jp }}\, , 
Miki WADATI $^{*}$
\vspace{5mm}\\
{\it $^{*}$Department of Physics, Graduate School of Science,}\\
{\it University of Tokyo,}\\
{\it Hongo 7-3-1, Bunkyo-ku, Tokyo 113-0033, Japan}
\vspace{3mm}\\
{\it $^{\ddag}$Department of Physics, Tokyo Institute of Technology,}\\
{\it Oh-okayama 2-12-1, Meguro-ku, Tokyo 152-8551, Japan}\\
}
\date{}
\maketitle

\begin{abstract}
We study the one-dimensional asymmetric simple exclusion 
process (ASEP) with open boundary conditions. 
Particles are injected and ejected at both boundaries. 
It is clarified that the steady state of the model is intimately
related to the Askey-Wilson polynomials.
The partition function and the $n$-point functions are obtained 
in the integral form with four boundary parameters.
The thermodynamic current is evaluated to confirm the 
conjectured phase diagram.

\vspace*{3mm}
\noindent
[Keywords: asymmetric simple exclusion process, 
exactly solvable model, Askey-Wilson polynomials]
\end{abstract}

\pagebreak

%%%%%%%%%%%%%%%%%%%%%%%%

\setcounter{equation}{0}
\section{Introduction}

In spite of its long history, 
statistical mechanics for nonequilibrium systems remains to be 
established. The central principle for the general theory 
is still lacking. Recently, a lot of analytical and numerical 
studies of specific nonequilibrium models have revealed their 
highly interesting behaviors. Among them, exactly 
solvable models are of great value because physical quantities and 
their relations are obtained explicitly \cite{SZ,Privman,Schuetz}. 

The one-dimensional asymmetric simple exclusion process (ASEP) is 
an exactly solvable stochastic process of many particles
\cite{Schuetz, Ligg, Ligg2, Sp}. 
Particles subject to the exclusion interaction perform
random walk with different rates in the right and the left 
direction.
Even after the relaxation to the steady state, 
the ASEP has nonvanishing particle current, 
maintained in a nonequilibrium situation.
The ASEP has become one of the standard models for 
the study of nonequilibrium statistical mechanics
because of its simplicity, rich behaviors and wide 
applicability. For instance, the ASEP is regarded as a 
primitive model of kinetics of biopolymerization\cite{MGPipkin}, 
traffic flow\cite{Schreck}, 
formation of shocks\cite{DJLS}. 
It is a discrete version of a hydrodynamic system obeying 
noisy Burgers equation\cite{BerGia97}. 
It also appears in a kind of sequence alignment problems 
in computational biology\cite{Bund}. 

In this article we study the ASEP with open boundary conditions. 
The model has in total five parameters $\alpha,\beta,\gamma,\delta$ and $q$.
The parameter $q$ is related to the asymmetry of the 
particle hopping; $\alpha$ (resp. $\delta$) and $\gamma$ 
(resp. $\beta$) represents the particle input (resp. output) rate 
at the left and the right boundary respectively.
The stationary properties of the model has attracted much attention. 
Among several reasons, one of them may be that it shows a boundary 
induced phase transition\cite{Krug91}; even the bulk properties of the 
system change drastically depending on the values of the 
boundary parameters.
For the case where particles hop only in one direction  
($q=\gamma=\delta=0$), the current and the density in the
thermodynamic limit was calculated in \cite{DEHP,SD1993}. 
In \cite{DEHP}, the authors introduced a novel algebraic 
formulation of the problem, which we call the matrix method 
in the following. 
This is a method of constructing the exact stationary measure 
of the model in terms of matrix products. It 
has become a standard techniques for investigating stationary 
properties of the ASEP and related models.
A lot of generalizations have been considered 
including discrete time models and multispecies models.

In \cite{DEHP}, for the continuous time setting, 
the algebraic formulation of the solution 
has been given for case with the full five parameters
although the computation of physical quantities in the thermodynamic 
limit was performed only for the case where $q=\gamma=\delta=0$.
In \cite{Sandow94} the five parameter case was treated.
The current in the thermodynamic limit was calculated and 
the phase diagram was obtained. The analysis, however, 
included some approximations.  In addition,
the computations of other quantities such as the density 
using the same direction of argument seemed difficult.
In \cite{SS99,BECE}, the case where $\gamma=\delta=0$ was studied.
The analysis of \cite{SS99} was based on the findings
that the stationary state of the ASEP for the case is 
related to the $q$-orthogonal polynomials called 
the Al-Salam-Chihara polynomials. This method has also admitted 
the computation of the density profile in \cite{SS2000}.

The objective of this paper is to investigate the most general case with 
the five parameters and further clarify 
the intimate relationship between the stationary state 
of the ASEP and the theory of $q$-orthogonal polynomials. 
We exactly calculate the partition function, 
the $n$-point functions and other physical quantities. 
Each quantity is written in an explicit integral form in terms of 
the weight function of the Askey-Wilson polynomials. 
The asymptotic behaviors in the thermodynamic limit are easily evaluated 
using analytical methods for integrals. 
As a result, we confirm the phase diagram for the current obtained 
in \cite{Sandow94}.  

The paper is organized as follows. 
In section \ref{sec:model}, the definition of the model and 
physical quantities of the ASEP are introduced. 
The matrix method is also explained. 
In section \ref{sec:AW}, properties of the Askey-Wilson polynomial are reviewed. 
The recurrence relation and the orthogonality relation play key roles 
in deriving the integral formula of the partition function. 
In section \ref{sec:repq<1}, two representations of the 
algebraic relations of matrices are given. 
One of them is related to the $q$-Hermite polynomial, which is the 
special case of the Askey-Wilson polynomial. 
It is simpler but the parameters are restricted. 
The other is related to the Askey-Wilson polynomial with full 
five parameters where no restriction is needed. 
The connection between the two representations is also mentioned. 
Section \ref{sec:repq=1} deals with the case where $q=1$. 
In section \ref{sec:Z_L}, integral formulae of the partition function are 
derived. In the thermodynamic limit, the asymptotic behaviors of the partition 
function and bulk quantities of the ASEP are examined.
There appears nonequilibrium phase transition. 
In section \ref{sec:nPointFunction}, using formulae in the $q$-calculus, 
integral formulae of the $n$-point functions are derived. 
The last section is devoted to the conclusion.

\setcounter{equation}{0}
\section{Model and Matrix Method}
\label{sec:model}

The one-dimensional ASEP is a model of randomly hopping 
particles with exclusion interaction on a one-dimensional lattice 
\cite{SZ, Privman, Schuetz, Ligg, Ligg2, Sp}. 
Let us denote the system size as $L$. 
Each site $i=1,\cdots,L$ of the lattice can afford at most only one particle; 
two or more particles cannot stay on the same site due to the hard-core 
exclusion interaction. A particle hops to the nearest right 
(resp. left) site with the probability $p_Rdt$ (resp. $p_Ldt$) 
during an infinitesimal time $dt$ if the target site is empty. 
By rescaling time, we can set $p_R=1$ and $p_L=q$. 
In the case of open boundaries, a particle enters the 
system at the left (resp. right) boundary with rate 
$\alpha$ (resp. $\delta$) if the site $1$ (resp. $L$) is empty, and is removed 
at the left (resp. right) boundary with rate $\gamma$ (resp. $\beta$) 
if the site $1$ (resp. $L$) is occupied. 
Figure \ref{asepfig} illustrates how the system evolves. 
Obviously, we can think of the reflection symmetry in this finite system: 
\begin{eqnarray}
&&\mathrm{site}\qquad i\leftrightarrow L-i+1, \qquad i=1,\cdots,L\nonumber\\
&&\mathrm{rate}\qquad q \leftrightarrow q^{-1},\ 
\alpha \leftrightarrow q^{-1}\delta,\ 
\beta \leftrightarrow q^{-1}\gamma,\ 
\gamma \leftrightarrow q^{-1}\beta,\ 
\delta \leftrightarrow q^{-1}\alpha .
\label{eqn:rs}
\end{eqnarray}
Therefore, 
the case where $q>1$ is transformed into the case where $q<1$ by the symmetry. 
In the following, we only assume the boundary parameters 
$\alpha, \beta, \gamma, \delta$ are strictly positive
unless otherwise stated.

The configuration of the system is labeled by 
$(\tau_1, \tau_2, \cdots ,\tau_L)$ where $\tau_i=0, 1$ is the
particle number at site $i$. 
Let $P(\tau_1, \tau_2,\cdots , \tau_L;t)$ denote the probability 
in the configuration $(\tau_1, \tau_2,\cdots , \tau_L)$ at time $t$. 
The time evolution of $P(\tau_1, \tau_2,\cdots , \tau_L;t)$ is 
described by the master equation, 
\begin{eqnarray}
&&
\frac{d}{dt}P(\tau_1,\tau_2,\cdots,\tau_L;t)= \nonumber\\
&&\quad
\delta_{\tau_1,1}[\alpha P(0,\tau_2,\cdots,\tau_L;t)
-\gamma P(1,\tau_2,\cdots,\tau_L;t)] \nonumber\\
&&\quad
+\delta_{\tau_1,0}[ \gamma P(1,\tau_2,\cdots,\tau_L;t)
-\alpha P(0,\tau_2,\cdots,\tau_L;t)]\nonumber\\
&&\quad
+\sum_{i=1}^{L-1}\biggl\{ 
\delta_{(\tau_i,\tau_{i+1})=(1,0)}
[qP(\cdots,\tau_i=0,\tau_{i+1}=1,\cdots;t)
-P(\cdots,\tau_i=1,\tau_{i+1}=0,\cdots;t)]\nonumber\\
&&\quad
+\delta_{(\tau_i,\tau_{i+1})=(0,1)}
[ P(\cdots,\tau_i=1,\tau_{i+1}=0,\cdots;t)
-qP(\cdots,\tau_i=0,\tau_{i+1}=1,\cdots;t)]
\biggr\}\nonumber\\
&&\quad
+\delta_{\tau_L,0}
[ \beta  P(\tau_1,\cdots,\tau_{L-1},1;t)
 -\delta P(\tau_1,\cdots,\tau_{L-1},0;t)]\nonumber\\
&&\quad
+\delta_{\tau_L,1}
[ \delta P(\tau_1,\cdots,\tau_{L-1},0;t)
 -\beta P(\tau_1,\cdots,\tau_{L-1},1;t)].
\end{eqnarray}
Hereafter, we focus on the stationary situation of the ASEP. 
According to the matrix method \cite{DEHP}, 
the stationary solution $P(\tau_1,\tau_2,\cdots,\tau_L)$ 
for the master equation is obtained 
in the form,
\begin{eqnarray}
P(\tau_1,\tau_2,\cdots,\tau_L)=
\frac{1}{Z_L}\llangle W\vert 
\prod_{i=1}^{L\atop\longrightarrow}(\tau_i\D+(1-\tau_i)\E) 
\vert V\rrangle.
\label{eqn:MPA}
\end{eqnarray}
Here $\D$ and $\E$ are matrices; $\D$ (resp. $\E$) 
corresponds to an occupation 
(resp. emptiness) of a particle. 
$\llangle W\vert$ and $\vert V\rrangle$ are vectors; $\llangle W\vert$ 
(resp. $\vert V\rrangle$) corresponds to the left (resp. right) boundary. 
The normalization constant $Z_L$ is given by
\begin{eqnarray}
Z_L=\llangle W\vert \C^L\vert V\rrangle, 
\label{eqn:defZ_L}
\end{eqnarray}
where 
\begin{eqnarray}
\C=\D+\E.
\end{eqnarray}
We call $Z_L$ the partition function. 
One can show that (\ref{eqn:MPA}) gives the exact steady 
state of the system if the matrices and the vectors satisfy
\cite{DEHP} 
\begin{eqnarray}
&&\D\E-q\E\D=\D+\E,
\label{eqn:relDE}\\
&&\llangle W\vert(\alpha \E-\gamma \D)=\llangle W\vert ,
\label{eqn:relW1}\\
&&(\beta \D-\delta \E)\vert V\rrangle =\vert V\rrangle.
\label{eqn:relV1}
\end{eqnarray}

Physical quantities are written in the form of matrix products. 
The average particle number at site $i$, $\langle \tau_i\rangle$, 
where the bracket means the average over the stationary 
probability distribution (\ref{eqn:MPA}), is written as 
\begin{eqnarray}
\langle \tau_i\rangle =
\frac{1}{Z_L}\llangle W\vert \C^{i-1}\D\C^{L-i}\vert V\rrangle,
\label{def:1Pfn}
\end{eqnarray}
and the two-point function $\langle \tau_i\tau_j\rangle$ as 
\begin{eqnarray}
\langle \tau_i\tau_j\rangle =
\frac{1}{Z_L}\llangle W\vert \C^{i-1}\D\C^{j-i-1}\D\C^{L-j}\vert V\rrangle.
\label{def:2Pfn}
\end{eqnarray}
The $n$-point functions are expressed similarly. 
The particle current through the bond between the neighboring sites 
from left to right, 
which is defined by 
$J=\langle \tau_i(1-\tau_{i+1})-q(1-\tau_i)\tau_{i+1}\rangle$, 
is simply given by
\begin{eqnarray}
J=\frac{Z_{L-1}}{Z_L}.
\label{def:J}
\end{eqnarray}
The site-dependence of $J$ is unnecessary since the current 
is uniform for the steady state. 
One can introduce the fugacity $\xi^2$ in $Z_L$ as
\begin{eqnarray}
Z_L(\xi^2)=\llangle W\vert (\xi^2\D+\E)^L\vert V\rrangle.
\label{eqn:Zfuga}
\end{eqnarray}
The square of fugacity in the definition is just for convenience. 
Then, the mean and the variance of the particle density in the 
whole system are calculated from the first and the second 
derivative of $Z_L(\xi^2)$ \cite{SS2000b}, 
\begin{eqnarray}
&&\langle \rho\rangle = 
\frac{1}{L}\xi^2\frac{\partial}{\partial\xi^2}
\log Z_L(\xi^2)\bigg\vert_{\xi^2=1},
\label{def:rho}\\
&&\langle \Delta\rho^2\rangle = 
\left(\frac{1}{L}\xi^2\frac{\partial}{\partial\xi^2}\right)^2
\log Z_L(\xi^2)\bigg\vert_{\xi^2=1}.
\label{def:rho^2}
\end{eqnarray}
As is familiar in equilibrium statistical physics, 
the bulk quantities are written in terms of the partition function.

\setcounter{equation}{0}
\section{Askey-Wilson Polynomial}
\label{sec:AW}

The Askey-Wilson polynomial is a $q$-orthogonal polynomial 
with four free parameters besides $q$. 
It resides on the top of the hierarchy of the one-variable 
$q$-orthogonal polynomial family in the Askey scheme\cite{AW,GR,Koekoek}. 
We list several properties of the Askey-Wilson polynomial 
needed in the following sections. For the purpose we introduce notations 
heavily used in the $q$-calculus. In this section, $\vert q\vert<1$ is assumed. 
The $q$-shifted factorial: 
\begin{eqnarray}
(a_1,a_2,\cdots,a_s;q)_n=\prod_{r=1}^s \prod_{k=0}^{n-1} (1-a_rq^k).
\end{eqnarray}
The basic hypergeometric function: 
\begin{eqnarray}
{}_r\phi_s\left[ {{a_1,\cdots ,a_r}\atop{b_1,\cdots ,b_s}};q,z\right]
=\sum_{k=0}^\infty \frac{(a_1,\cdots,a_r;q)_k}{(b_1,\cdots,b_s,q;q)_k}
((-1)^k q^{k(k-1)/2})^{1+s-r} z^k.
\end{eqnarray}
Then, the Askey-Wilson polynomial $P_n(x)=P_n(x;a,b,c,d\vert q)$
is explicitly defined by 
\begin{eqnarray}
P_n(x)=
a^{-n}(ab,ac,ad;q)_n\ 
{}_4\phi_3\left[ {{q^{-n},q^{n-1}abcd,ae^{i\theta},ae^{-i\theta}}
	\atop{ab,ac,ad}};q,q \right] , 
\label{eqn:defAW}
\end{eqnarray}
with $x=\cos\theta$ for $n\in \Z_+:=\{0,1,2,\cdots\}$. 
It satisfies the three-term recurrence relation, 
\begin{eqnarray}
A_nP_{n+1}(x)+B_nP_n(x)+C_nP_{n-1}(x)=2xP_n(x), 
\label{eqn:recAW}
\end{eqnarray}
with $P_0(x)=1$ and $P_{-1}(x)=0$, 
where
\begin{align}
A_n&=
\frac{1-q^{n-1}abcd}{(1-q^{2n-1}abcd)(1-q^{2n}abcd)}, 
\label{eqn:A_n}\\
B_n&=
\frac{q^{n-1}}{(1-q^{2n-2}abcd)(1-q^{2n}abcd)} 
[(1+q^{2n-1}abcd)(qs+abcds')-q^{n-1}(1+q)abcd(s+qs')], 
\label{eqn:B_n}\\
C_n&=
\frac{(1-q^n)(1-q^{n-1}ab)(1-q^{n-1}ac)(1-q^{n-1}ad)(1-q^{n-1}bc)
(1-q^{n-1}bd)(1-q^{n-1}cd)}
{(1-q^{2n-2}abcd)(1-q^{2n-1}abcd)}, 
\label{eqn:C_n}
\end{align}
and
\begin{eqnarray}
s=a+b+c+d, \qquad s'=a^{-1}+b^{-1}+c^{-1}+d^{-1}. 
\end{eqnarray}
From this relation, we see that $P_n(x;a,b,c,d\vert q)$ is symmetric 
with respect to $a$, $b$, $c$ and $d$. 
The orthogonality relation depends on the parameter region. 
If $\vert a\vert, \vert b\vert, \vert c\vert, \vert d\vert <1$, 
the orthogonality reads 
\begin{eqnarray}
\int_0^{\pi}\frac{d\theta}{2\pi}
w(\cos\theta)P_m(\cos\theta)P_n(\cos\theta)
=h_n\delta_{mn}, 
\label{eqn:orthoAW}
\end{eqnarray}
where
\begin{eqnarray}
&&w(\cos\theta)=
\frac{(e^{2i\theta},e^{-2i\theta};q)_\infty}
{(ae^{i\theta},ae^{-i\theta},be^{i\theta},be^{-i\theta},
ce^{i\theta},ce^{-i\theta},de^{i\theta},de^{-i\theta};q)_\infty}, \\
&&\frac{h_n}{h_0}=
\frac{(1-q^{n-1}abcd)(q,ab,ac,ad,bc,bd,cd;q)_n}
{(1-q^{2n-1}abcd)(abcd;q)_n} ,\\
&&h_0=
\frac{(abcd;q)_\infty}{(q,ab,ac,ad,bc,bd,cd;q)_\infty} .
\end{eqnarray}
If we rewrite (\ref{eqn:orthoAW}) with $z=e^{i\theta}$, we have 
\begin{eqnarray}
\oint_C \frac{dz}{4\pi iz} w\left(\frac{z+z^{-1}}{2}\right)
P_m\left(\frac{z+z^{-1}}{2}\right)P_n\left(\frac{z+z^{-1}}{2}\right)
=h_n\delta_{mn}, 
\label{eqn:orthoointAW}
\end{eqnarray}
where the integral contour $C$ is a closed path which 
encloses the poles at $z=aq^k$, $bq^k$, $cq^k$, $dq^k$ $(k\in \Z_+)$ 
and excludes the poles at $z=(aq^k)^{-1}$, $(bq^k)^{-1}$, $(cq^k)^{-1}$, 
$(dq^k)^{-1}$ $(k\in \Z_+)$. 
In the other parameter region, the orthogonality is continued analytically. 
Putting $m=n=0$, we have the celebrated Askey-Wilson integral, 
\begin{eqnarray}
\oint_C \frac{dz}{4\pi iz}
\frac{(z^2,z^{-2};q)_\infty}{(az,a/z,bz,b/z,cz,c/z,dz,d/z;q)_\infty}
=\frac{(abcd;q)_\infty}{(q,ab,ac,ad,bc,bd,cd;q)_\infty}.
\end{eqnarray}
For the case of $a=b=c=d=0$, the Askey-Wilson polynomial is reduced to 
the $q$-Hermite polynomial, 
\begin{eqnarray}
H_n(x\vert q)=P_n(x;0,0,0,0\vert q).
\end{eqnarray}
Its recurrence and orthogonality relation are those with $a=b=c=d=0$ in the 
above formulae for $P_n(x;a,b,c,d\vert q)$. 
In what follows, 
we sometimes write briefly the corresponding weight function or coefficients 
as $f^{(0)}:=f\vert_{a=b=c=d=0}$. 
We introduce the $q$-binomial, 
\begin{eqnarray}
F_n(x,y)=
\sum_{k=0}^n\frac{(q;q)_n}{(q;q)_k(q;q)_{n-k}}x^{n-k}y^k ,
\label{eqn:defF}
\end{eqnarray}
which satisfies the following recurrence relation, 
\begin{eqnarray}
F_{n+1}(x,y)-(x+y)F_n(x,y)+(1-q^n)xyF_{n-1}(x,y)=0 .
\label{eqn:recF}
\end{eqnarray}
with $F_0(x,y)=1$ and $F_{-1}(x,y)=0$. 
The $q$-Hermite polynomial is the special case of it: 
\begin{eqnarray}
H_n(\cos\theta\vert q)=F_n(e^{i\theta},e^{-i\theta}).
\label{eqn:H&F}
\end{eqnarray} 
As to $F_n(x,y)$, the $q$-Mehler formula is important: 
\begin{eqnarray}
\sum_{n=0}^\infty \frac{F_n(v,w)F_n(x,y)}{(q;q)_n}\xi ^n = 
\frac{(vwxy\xi ^2;q)_\infty}{(vx\xi ,vy\xi ,wx\xi ,wy\xi ;q)_\infty},
\label{eqn:qMehler}
\end{eqnarray}
for $\vert vx\xi\vert$, $\vert vy\xi\vert$, $\vert wx\xi\vert$, 
$\vert wy\xi\vert <1$. 
Combining (\ref{eqn:recF}) with this, we obtain another kind of bilinear 
function: 
\begin{eqnarray}
\sum_{n=0}^\infty \frac{F_n(v,w)F_{n+1}(x,y)}{(q;q)_n}\xi ^n = 
\frac{x+y-(v+w)xy\xi}{1-vwxy\xi^2}
\frac{(vwxy\xi ^2;q)_\infty}{(vx\xi ,vy\xi ,wx\xi ,wy\xi ;q)_\infty},
\label{eqn:qMehler2}
\end{eqnarray}
for $\vert vx\xi\vert$, $\vert vy\xi\vert$, $\vert wx\xi\vert$, 
$\vert wy\xi\vert <1$. 

When $abcd\neq 0$, the symmetry relation for ${}_4\phi_3$,
\begin{equation}
 {}_4\phi_3\left[ {{q^{-n},a,b,c}\atop{d,e,f}};q,q \right] 
 =
 {}_4\phi_3\left[     {{q^n,a^{-1},b^{-1},c^{-1}}
                  \atop{d^{-1},e^{-1},f^{-1}}};q^{-1},q^{-1} \right],
\end{equation}
implies the symmetry relation for the Askey-Wilson polynomials, 
\begin{eqnarray}
P_n(x;a,b,c,d\vert q)=
(-abcd)^n q^{3n(n-1)/2} P_n(x;a\inv,b\inv,c\inv,d\inv\vert q\inv).
\end{eqnarray}
From this symmetry, it is enough to study the case where $|q|<1$.
But when $abcd=0$, this relation no longer holds and one
has to consider the case where $q>1$ separately \cite{Askey87,IM94}.
For instance suppose that only one of $a,b,c,d$ is zero.
Without loss of generality, we can set $d=0$ because of the 
symmetry in $a,b,c,d$. 
The recurrence relation (\ref{eqn:recAW}) remains the same 
(with $d$ set to zero), but the orthogonality relation 
(\ref{eqn:orthoAW}) should be modified. 
$P_n(x)$'s are now orthogonal on the imaginary axis;
\begin{align}
\int_{-\infty}^\infty \!\!\!du
\ \ \widetilde{w}(\sinh u) P_m(i\sinh u)P_n(i\sinh u)
=\widetilde{h}_n\delta_{mn}, 
\label{eqn:ortho-1AW}
\end{align}
where
\begin{align}
&\widetilde{w}(\sinh u)=\frac{1}{\log q}
\frac{(iaq^{-1}e^u,-iaq^{-1}e^{-u},
ibq^{-1}e^u,-ibq^{-1}e^{-u},
icq^{-1}e^u,-icq^{-1}e^{-u};\Q)_\infty}
{(-q^{-1}e^{2u},-q^{-1}e^{-2u};q^{-1})_\infty},
\label{eqn:-1w}\\
&\widetilde{h}_n=
(q,ab,ac,bc;q)_n (q^{-1},abq^{-1},acq^{-1},bcq^{-1};q^{-1})_\infty .
\label{eqn:-1h_n}
\end{align}
The formulae when more than one of $a,b,c,d$ are zero are similar.
For details about the case where $q>1$,
we refer the reader to \cite{Askey87,IM94}.
The list of formulae exploited in the following sections is completed.

\setcounter{equation}{0}
\section{Representation: the case where $q\neq 1$}
\label{sec:repq<1}

In the subsequent sections we give two infinite-dimensional representations of 
matrices and vectors. One is related to the $q$-Hermite polynomials 
and the other to the Askey-Wilson polynomials. 
The representations are useful since the $q$-orthogonal 
polynomial forms a complete set of eigenvectors of the matrix $\C$, 
which enters in the matrix expression of $Z_L$ (\ref{eqn:defZ_L}). 
This leads to an integral formula of it. 
In the following the parameters are fixed as 
\begin{eqnarray}
a=\kappa_{\alpha,\gamma}^+,\ b=\kappa_{\beta,\delta}^+,\ 
c=\kappa_{\alpha,\gamma}^-,\ d=\kappa_{\beta,\delta}^-,
\label{abcd}
\end{eqnarray}
where 
\begin{eqnarray}
&&\kappa_{\alpha,\gamma}^\pm =
\frac{1}{2\alpha}
\left[ (1-q-\alpha+\gamma)\pm
\sqrt{(1-q-\alpha+\gamma)^2+4\alpha\gamma}\right] ,\\
&&\kappa_{\beta,\delta}^\pm =
\frac{1}{2\beta}
\left[ (1-q-\beta+\delta)\pm
\sqrt{(1-q-\beta+\delta)^2+4\beta\delta}\right] .
\end{eqnarray}
We also fix $q$ as $0<q<1$; the $q>1$ case  
is simply considered from the reflection symmetry (\ref{eqn:rs}), 
or equivalently, 
\begin{eqnarray}
a,b,c,d,q\ \leftrightarrow \ b\inv ,a\inv ,d\inv ,c\inv ,q\inv .
\label{eqn:rs1}
\end{eqnarray}
Since $\alpha,\beta,\gamma,\delta$ are strictly positive, 
the new parameters satisfy $a>0$, $b>0$, $-1<c<0$ and $-1<d<0$ if $0<q<1$. 
The bra and ket notation are used for row and column vectors, respectively. 
In place of matrices $\D$ and $\E$, we consider $\dd$ and $\ee$, 
\begin{eqnarray}
&&\D =\frac{1}{1-q}(\1+\dd),\qquad \E =\frac{1}{1-q}(\1+\ee),
\label{eqn:defde}
\\
&&\C=\D+\E=\frac{1}{1-q}(2\1+\dd+\ee).
\label{eqn:defC}
\end{eqnarray}
For them the relations (\ref{eqn:relDE}), (\ref{eqn:relW1}) and 
(\ref{eqn:relV1}) are replaced by 
\begin{eqnarray}
&&\dd\ee-q\ee\dd = (1-q)\1 ,
\label{eqn:rel-de}\\
&&\llangle W\vert [\ee-(a+c)\1+ac\dd]=0,
\label{eqn:rel-W}\\
&&[\dd-(b+d)\1+bd\ee] \vert V\rrangle =0.
\label{eqn:rel-V}
\end{eqnarray}

\subsection{Representation related to the $q$-Hermite polynomials}

First, we give a representation related to the $q$-Hermite polynomials. 
One can check the following matrices and vectors satisfy the relation 
(\ref{eqn:rel-de}), (\ref{eqn:rel-W}) and (\ref{eqn:rel-V}). 
\begin{eqnarray}
\dd=\left[
\begin{array}{cccc}
0 		& \sqrt{1-q} 	& 0	 	& \cdots\\
0	 	& 0	 	& \sqrt{1-q^2} 	& {}\\
0 		& 0	 	& 0	 	& \ddots\\
\vdots 		& {} 		& \ddots	& \ddots
\end{array}
\right] ,
&&\quad 
\ee=\left[
\begin{array}{cccc}
0	 	& 0	 	& 0 		& \cdots\\
\sqrt{1-q} 	& 0	 	& 0	 	& {}\\
0 		& \sqrt{1-q^2} 	& 0	 	& \ddots\\
\vdots 		& {}		& \ddots	& \ddots
\end{array}
\right],
\label{eqn:repde1}
\end{eqnarray}
\begin{eqnarray}
&&\llangle W\vert =
\nu_{a,c}\( F_0(a,c)\big/\sqrt{(q;q)_0},F_1(a,c)\big/\sqrt{(q;q)_1},
\cdots \) ,
\nonumber\\
&&\vert V\rrangle =
\nu_{b,d}\( F_0(b,d)\big/\sqrt{(q;q)_0},F_1(b,d)\big/\sqrt{(q;q)_1},
\cdots \) ^T,
\label{eqn:repWV1}
\end{eqnarray}
where $\nu_{a,c}=(ac;q)_\infty^{-1} (q;q)_\infty^{-1/2}$. 
We regard (\ref{eqn:rel-de}) as the $q$-deformed commutation relation of 
boson operators\cite{Bie,Mac,Kul-Dam}, and (\ref{eqn:repde1}) and (\ref{eqn:repWV1}) 
as the $q$-deformed Fock representation\cite{BECE}. 
Note that matrices $\dd$ and $\ee$ depend only on $q$, and vectors 
$\langle W\vert$ and $\vert V\rangle$ depend on the boundary 
parameters as well. 
In this representation, a vector defined by 
\begin{eqnarray}
&&\vert h(x)\rangle =(h_0(x),h_1(x),h_2(x),\cdots)^T,
\end{eqnarray}
where
\begin{eqnarray}
h_n(x)=\sqrt{\frac{(q;q)_\infty}{(q;q)_n}}H_n(x\vert q),
\label{eqn:defh}
\end{eqnarray}
is eigenvector of the matrix $\dd+\ee$ with the eigenvalue $2x$: 
\begin{eqnarray}
(\dd +\ee )\vert h(x)\rangle=2x\vert h(x)\rangle .
\end{eqnarray}
Each element of this equation corresponds to the recurrence relation of 
the $q$-Hermite polynomials. 
The transposed equation also holds for $\langle h(x)\vert :=
\vert h(x)\rangle^T$ 
because $\dd+\ee$ is a symmetric matrix. 
The orthogonality relation is rewritten as 
\begin{eqnarray}
\1=\int_0^{\pi}\frac{d\theta}{2\pi} w^{(0)}(\cos\theta)\ 
\vert h(\cos\theta)\rangle \langle h(\cos\theta)\vert .
\label{eqn:orthoh}
\end{eqnarray}
Therefore, the eigenvector of $\dd+\ee$, or at the same time the eigenvector 
of $\C$ from (\ref{eqn:defC}), 
$\{\vert h(\cos\theta)\rangle ;\theta\in[0,\pi]\}$ forms 
a complete set of vectors in this representation space. 
Furthermore, 
from (\ref{eqn:qMehler2}) with (\ref{eqn:H&F}) we obtain the following 
form factor, 
\begin{eqnarray}
\langle h(\cos\theta)\vert \Lambda(\xi) \dd \vert h(\cos\phi)\rangle 
=\frac{(q,\xi^2q;q)_\infty(2\cos\phi-2\xi\cos\theta)}
{(\xi e^{i(\theta +\phi)},\xi e^{-i(\theta +\phi)},
\xi e^{i(\theta -\phi)},\xi e^{-i(\theta -\phi)};q)_\infty}. 
\label{eqn:hdh-h}
\end{eqnarray}
Here a regularization matrix $\Lambda(\xi)=\mathrm{diag}(1,\xi,\xi^2,\cdots)$ 
is introduced. 

In fact, this representation is valid only when 
$\vert a\vert,\vert b\vert,\vert c\vert,\vert d\vert<1$, 
otherwise, the boundary vectors are ill-defined.
For example, if $a>1$, $\langle W\vert h(\cos\theta)\rangle$ becomes 
a divergent series. See the convergence condition for the $q$-Mehler formula 
(\ref{eqn:qMehler}). 
In the next subsection, we present another representation without such restriction.

\subsection{Representation related to the Askey-Wilson polynomial}
\label{sec:repAWq<1}

We consider a representation related to the Askey-Wilson polynomials with 
full five parameters. Here $a,b,c,d$ are not restricted. 
Complicated apparently, the followings certainly solve the equations 
(\ref{eqn:rel-de}), (\ref{eqn:rel-W}) and (\ref{eqn:rel-V}):
\begin{eqnarray}
\dd=\left[
\begin{array}{cccc}
d_0^\natural 	& d_0^\sharp 	& 0	 	& \cdots\\
d_0^\flat 	& d_1^\natural 	& d_1^\sharp 	& {}\\
0 		& d_1^\flat 	& d_2^\natural 	& \ddots\\
\vdots 		& {} 		& \ddots	& \ddots
\end{array}
\right] ,
&&\qquad 
\ee=\left[
\begin{array}{cccc}
e_0^\natural 	& e_0^\sharp 	& 0 		& \cdots\\
e_0^\flat 	& e_1^\natural 	& e_1^\sharp 	& {}\\
0 		& e_1^\flat 	& e_2^\natural 	& \ddots\\
\vdots 		& {}		& \ddots	& \ddots
\end{array}
\right],
\label{eqn:repde2}
\end{eqnarray}
\begin{eqnarray}
&&\llangle W\vert =h_0^{1/2}(1,0,0,\cdots ), \qquad
\vert V\rrangle =h_0^{1/2}(1,0,0,\cdots )^T, 
\label{eqn:repWV2}
\end{eqnarray}
where 
\begin{eqnarray}
d_n^\natural &=&
\frac{q^{n-1}}{(1-q^{2n-2}abcd)(1-q^{2n}abcd)}\nonumber\\
&&\times[
bd(a+c)+(b+d)q-abcd(b+d)q^{n-1}-\{ bd(a+c)+abcd(b+d)\} q^n\nonumber\\
&&-bd(a+c)q^{n+1}+ab^2 cd^2(a+c) q^{2n-1}+abcd(b+d)q^{2n} ] ,\\
e_n^\natural &=&
\frac{q^{n-1}}{(1-q^{2n-2}abcd)(1-q^{2n}abcd)}\nonumber\\
&&\times[
ac(b+d)+(a+c)q-abcd(a+c)q^{n-1}-\{ ac(b+d)+abcd(a+c)\} q^n\nonumber\\
&&-ac(b+d)q^{n+1}+a^2 bc^2 d(b+d) q^{2n-1}+abcd(a+c)q^{2n} ] ,
\label{eqn:elements}
\end{eqnarray}
\begin{eqnarray}
&&d_n^\sharp =
\frac{1}{1-q^nac}\mathcal{A}_n ,\qquad
e_n^\sharp =
-\frac{q^nac}{1-q^nac}\mathcal{A}_n ,\\
&&d_n^\flat =
-\frac{q^nbd}{1-q^nbd}\mathcal{A}_n ,\qquad
e_n^\flat =
\frac{1}{1-q^nbd}\mathcal{A}_n ,
\end{eqnarray}
and 
\begin{align}
\mathcal{A}_n=
\left[
\frac{(1-q^{n-1}abcd)(1-q^{n+1})
(1-q^nab)(1-q^nac)(1-q^nad)(1-q^nbc)(1-q^nbd)(1-q^ncd)}
{(1-q^{2n-1}abcd)(1-q^{2n}abcd)^2(1-q^{2n+1}abcd)}
\right]^{1/2} .
\end{align}
As before, a vector defined by 
\begin{equation}
\vert p(x)\rangle = (p_0(x),p_1(x),p_2(x),\cdots)^T,
\end{equation}
where 
\begin{eqnarray}
p_n(x)=
\sqrt{\frac{h_0}{h_n}}P_n(x;a,b,c,d\vert q),
\label{eqn:defp}
\end{eqnarray}
becomes the eigenvector of the matrix $\dd+\ee$ 
with the eigenvalue $2x$: 
\begin{eqnarray}
(\dd +\ee )\vert p(x)\rangle=2x\vert p(x)\rangle .
\label{eqn:eigenp}
\end{eqnarray}
Each element of this equation corresponds to the recurrence relation of 
the Askey-Wilson polynomials (\ref{eqn:recAW}). 
The transposed equation also holds for $\langle p(x)\vert := 
\vert p(x)\rangle^T$ 
because $\dd+\ee$ is a symmetric matrix. 
The orthogonality relation (\ref{eqn:orthoAW}) is rewritten as 
\begin{eqnarray}
\1=h_0^{-1}\oint_C \frac{dz}{4\pi iz} w\((z+z^{-1})/2\)\ 
\vert p\((z+z^{-1})/2\)\rangle \langle p\((z+z^{-1})/2\)\vert ,
\label{eqn:orthop}
\end{eqnarray}
where $C$ is as described below (\ref{eqn:orthoointAW}). 
The reflection (\ref{eqn:rs1}) corresponds to just taking transpositions 
in the representation. 
If we set $c=d=0$, or equivalently $\gamma=\delta=0$, 
we find the representation is reduced to the one related 
to the Al-Salam-Chihara polynomials\cite{SS99}. 

The relation between the representation here and the previous one is 
almost obvious. They are equivalent if 
$\vert a\vert,\vert b\vert,\vert c\vert,\vert d\vert<1$. 
The connection matrix comes from the transformation, 
\begin{eqnarray}
\frac{h_0^{-1/2}}
{(ae^{i\theta},ae^{-i\theta},ce^{i\theta},ce^{-i\theta};q)_\infty}
\vert p(\cos\theta)\rangle=
G^{(a,c)}_{(b,d)} \ \vert h(\cos\theta)\rangle .
\label{eqn:connection}
\end{eqnarray}
We can calculate the elements as 
\begin{eqnarray}
\left[ G^{(a,c)}_{(b,d)}\right]_{n,k}=
\frac{(ab,ac,ad;q)_n}
{a^n(ac;q)_\infty \sqrt{h_n(q;q)_k(q;q)_\infty}}
\sum_{j=0}^n 
\frac{(q^{-n},q^{n-1}abcd;q)_j q^j}
{(q,ab,ad;q)_j}
F_k(aq^j,c). 
\end{eqnarray}
The inverse matrix is 
\begin{eqnarray}
\left[ G^{(a,c)}_{(b,d)}\right] ^{-1} =
\left[ G^{(b,d)}_{(a,c)}\right]^T.
\end{eqnarray}
To prove this, compare the two orthogonality relations (\ref{eqn:orthoh}) 
and (\ref{eqn:orthop}). 
Replacing $(a,c)\leftrightarrow(b,d)$ and transposing in 
(\ref{eqn:connection}), we have 
\begin{eqnarray}
\frac{h_0^{-1/2}}
{(be^{i\theta},be^{-i\theta},de^{i\theta},de^{-i\theta};q)_\infty}
\langle p(\cos\theta)\vert=
\langle h(\cos\theta)\vert \left[ G^{(b,d)}_{(a,c)}\right]^T .
\end{eqnarray}
Then, 
\begin{eqnarray}
\1&=&h_0^{-1}\int_0^{\pi}\frac{d\theta}{2\pi} w(\cos\theta)\ 
\vert p(\cos\theta)\rangle \langle p(\cos\theta)\vert \nonumber\\
&=&\int_0^{\pi}\frac{d\theta}{2\pi} w^{(0)}(\cos\theta)\ 
G^{(a,c)}_{(b,d)}\ \vert h(\cos\theta)\rangle 
\langle h(\cos\theta)\vert \left[ G^{(b,d)}_{(a,c)}\right]^T \nonumber\\
&=&G^{(a,c)}_{(b,d)}\left[ G^{(b,d)}_{(a,c)}\right]^T. \nonumber
\end{eqnarray}
This matrix actually connects the two representations since, for example, 
$\langle W\vert^{\mathrm{Hermite}}=
\langle W\vert^{\mathrm{AW}}G^{(a,c)}_{(b,d)}$. 

Looking at the form factor in the $q$-Hermite representation (\ref{eqn:hdh-h}),
a nontrivial formula results in the Askey-Wilson representation;
\begin{align}
&\langle p(\cos\theta)\vert \Theta(\xi) \dd \vert p(\cos\phi)\rangle 
=\nonumber\\
&\qquad
\frac{(2\cos\phi-2\xi\cos\theta)(q,\xi^2q,
ae^{i\phi},ae^{-i\phi},ce^{i\phi},ce^{-i\phi},
be^{i\theta},be^{-i\theta},de^{i\theta},de^{-i\theta};q)_\infty}
{(\xi e^{i(\theta+\phi)},\xi e^{i(\theta-\phi)},\xi e^{-i(\theta+\phi)},
\xi e^{-i(\theta-\phi)};q)_\infty},
\label{eqn:formfactor}
\end{align}
where 
\begin{eqnarray}
\Theta(\xi)=
G^{(a,c)}_{(b,d)} \Lambda(\xi)  \left[ G^{(a,c)}_{(b,d)}\right] ^{-1}.
\end{eqnarray}
The matrix elements are explicitly calculated as 
\begin{align}
\left[ \Theta(\xi)\right]_{n,k}&=
\frac{1}{\sqrt{h_nh_k}}
\frac{(abcd\xi^2;q)_\infty (ab,ac,ad;q)_n (ab,bc,bd;q)_k}
{(q,ab\xi,ac,ad\xi,bc\xi,bd,cd\xi;q)_\infty}
a^{-n}b^{-k} \nonumber\\
&\times
\sum_{j=0}^n
\frac{(q^{-n},q^{n-1}abcd,ab\xi,ad\xi;q)_j q^j}
{(q,ab,ad,abcd\xi^2;q)_j}
{}_4\phi_3\left[ {{q^{-k},q^{k-1}abcd,ab\xi q^j,bc\xi}
\atop{ab,bc,abcd\xi q^j}};q,q\right].
\end{align}

A remark on a special case is in order. 
When we specialize to the case where $ab=q^{1-N}$ ($N=1,2,\cdots$), 
the infinite-dimensional representation can be reduced to an 
$N$-dimensional one. The same is true for other pairwise products of 
$a,b,c,d$ because of the symmetry in $a,b,c,d$. 
The matrices can be taken from the first $N\times N$ entry of 
those of the infinite-dimensional matrices, and similarly the first $N$ entry 
for the vectors: 
\begin{eqnarray}
\dd^{(N)}=\left[
\begin{array}{cccc}
d_0^\natural 	& d_0^\sharp 	& 0	 	& \cdots\\
d_0^\flat 	& d_1^\natural 	& \ddots 	& {}\\
0 		& \ddots	& \ddots 	& d_{N-2}^\sharp\\
\vdots 		& {} 		& d_{N-2}^\flat	& d_{N-1}^\natural
\end{array}
\right] ,
&&\qquad 
\ee^{(N)}=\left[
\begin{array}{cccc}
e_0^\natural 	& e_0^\sharp 	& 0	 	& \cdots\\
e_0^\flat 	& e_1^\natural 	& \ddots 	& {}\\
0 		& \ddots	& \ddots 	& e_{N-2}^\sharp\\
\vdots 		& {} 		& e_{N-2}^\flat	& e_{N-1}^\natural
\end{array}
\right],
\label{eqn:repde4}
\end{eqnarray}
\begin{eqnarray}
&&\llangle W^{(N)}\vert =h_0^{1/2}(1,0,\cdots ,0), \qquad
\vert V^{(N)}\rrangle =h_0^{1/2}(1,0,\cdots ,0)^T. 
\label{eqn:repWV4}
\end{eqnarray}
In this case, the corresponding Askey-Wilson polynomials are called 
the $q$-Racah polynomials. 
The orthogonality relation is rewritten in $N$ residue sum from the integral form. 
This special case may be useful for trial calculations of many quantities of the ASEP 
since a generally hard task of integration is replaced by finite summation. 
In fact, this finite-dimensional representation is equivalent to the one 
in \cite{Ess-Ritt96,Mall-Sand97}. 
The similarity transformation is realized by an $N\times N$ matrix $U$: 
\begin{eqnarray}
U_{nk}=
\frac{(ab,ac,ad;q)_n (ab,bc;q)_k}{a^n (-b)^k q^{k(k-1)/2} \sqrt{h_n}}
{}_3\phi_2\left[ {{q^{-n},q^{n-1}abcd,abq^k}
	\atop{ab,ad}};q,q \right] ,
\end{eqnarray}
for $n,k=0,\cdots,N-1$. 
Then, 
\begin{eqnarray}
&&U^{-1}\dd^{(N)} U=\left[
\begin{array}{ccccc}
b 	& 0 	& 0	 	& \cdots 	& \cdots\\
0 	& bq 	& \ddots 	& {}		& {}\\
0 	& 0	& bq^2		& \ddots 	& {}\\
\vdots 	& {} 	& \ddots	& \ddots	& 0\\
{}	& {}	& {}		& 0		& bq^{N-1}
\end{array}
\right] ,\\
&&
U^{-1} \ee^{(N)} U=\left[
\begin{array}{ccccc}
b^{-1} 	& 0 	& 0	 	& \cdots 	& \cdots\\
1 	& (bq)^{-1} 	& \ddots 	& {}		& {}\\
0 	& 1	& (bq^2)^{-1}		& \ddots 	& {}\\
\vdots 	& {} 	& \ddots	& \ddots	& 0\\
{}	& {}	& {}		& 1		& (bq^{N-1})^{-1}
\end{array}
\right],
\end{eqnarray}
\begin{eqnarray}
\llangle W^{(N)}\vert U &=&
(1,U_{01},\cdots ,U_{0,N-1}), \\
U^{-1}\vert V^{(N)}\rrangle &=&
\((U^{-1})_{00},(U^{-1})_{10},\cdots ,(U^{-1})_{N-1,0}\)^T. 
\end{eqnarray}

\setcounter{equation}{0}
\section{Representation: the case where $q=1$}
\label{sec:repq=1}

In this section a representation for $q=1$ is given. 
We consider matrices $\D$ and $\E$ instead of $\dd$ and $\ee$. 
Taking the limit $q\to1$ in the representation in section \ref{sec:repAWq<1} 
yields 
\begin{eqnarray}
\D=\left[
\begin{array}{cccc}
\DD_0^\natural 	& \DD_0^\sharp 	& 0	 	& \cdots\\
\DD_0^\flat 	& \DD_1^\natural& \DD_1^\sharp 	& {}\\
0 		& \DD_1^\flat 	& \DD_2^\natural 	& \ddots\\
\vdots 		& {} 		& \ddots	& \ddots
\end{array}
\right] ,
&&\qquad 
\E=\left[
\begin{array}{cccc}
\EE_0^\natural 	& \EE_0^\sharp 	& 0 		& \cdots\\
\EE_0^\flat 	& \EE_1^\natural& \EE_1^\sharp 	& {}\\
0 		& \EE_1^\flat 	& \EE_2^\natural& \ddots\\
\vdots 		& {}		& \ddots	& \ddots
\end{array}
\right],
\label{eqn:repDE3}
\end{eqnarray}
\begin{eqnarray}
&&\llangle W\vert =(1,0,0,\cdots ), \qquad
\vert V\rrangle =(1,0,0,\cdots )^T, 
\label{eqn:repWV3}
\end{eqnarray}
where 
\begin{eqnarray}
&&\DD^\natural_n=
\frac{\alpha+\delta+n(\alpha\beta+2\alpha\delta+\gamma\delta)}
{(\alpha+\gamma)(\beta+\delta)},
\qquad
\EE^\natural_n=
\frac{\beta+\gamma+n(\alpha\beta+2\beta\gamma+\gamma\delta)}
{(\alpha+\gamma)(\beta+\delta)},
\nonumber\\
&&\DD^\sharp_n=
\frac{\alpha}{\alpha+\gamma}
\left[(n+1)(\lambda+n+1)\right]^{1/2},
\qquad
\EE^\sharp_n=
\frac{\gamma}{\alpha+\gamma}
\left[(n+1)(\lambda+n+1)\right]^{1/2},
\nonumber\\
&&\DD^\flat_n=
\frac{\delta}{\beta+\delta}
\left[(n+1)(\lambda+n+1)\right]^{1/2},
\qquad
\EE^\flat_n=
\frac{\beta}{\beta+\delta}
\left[(n+1)(\lambda+n+1)\right]^{1/2},
\end{eqnarray}
with 
\begin{eqnarray}
\lambda=\frac{\alpha+\beta+\gamma+\delta}{(\alpha+\gamma)(\beta+\delta)}-1. 
\end{eqnarray}
The eigenvector of $\C$ is 
\begin{eqnarray}
\vert \ell (x)\rangle =(\ell_0(x),\ell_1(x),\cdots)^T, 
\end{eqnarray}
where the elements are 
\begin{eqnarray}
\ell_n(x)=(-1)^n
\left[\frac{n!\Gamma(\lambda+1)}
{\Gamma(\lambda+n+1)}\right]^{1/2}L^{(\lambda)}_{n}(x).
\end{eqnarray}
The orthogonal polynomials in this case are the Laguerre polynomials 
$L_n^{(\lambda)}(x)$. 
The recurrence relation of $L_n^{(\lambda)}(x)$ is 
\begin{eqnarray}
(n+1)L^{(\lambda)}_{n+1}(x)-(2n+\lambda+1-x)L^{(\lambda)}_{n}(x)+
(n+\lambda)L^{(\lambda)}_{n-1}(x)=0,
\end{eqnarray}
with $L^{(\lambda)}_{0}(x)=1$ and $L^{(\lambda)}_{-1}(x)=0$. 
In our notation it is rewritten as 
\begin{eqnarray}
\C\vert \ell (x)\rangle =x\vert \ell(x)\rangle .
\end{eqnarray}
The orthogonality relation reads 
\begin{eqnarray}
\int_0^\infty \!\! dx\ \ x^{\lambda}e^{-x} 
L^{(\lambda)}_{m}(x)L^{(\lambda)}_{n}(x)
=\frac{\Gamma(n+\lambda+1)}{n!}\delta_{mn},
\end{eqnarray}
for $\lambda>-1$. 
It is rewritten as 
\begin{eqnarray}
\1=\mu^{-1}\int_0^\infty \!\! dx\ \ x^{\lambda}e^{-x} 
\vert \ell(x)\rangle \langle \ell(x)\vert ,
\end{eqnarray}
where $\mu=\Gamma(\lambda+1)$. 
We can obtain the form factor as 
\begin{eqnarray}
\langle \ell(x)\vert \D \vert \ell(y)\rangle=
\mu^{-1}\left[
(c_1+c_2x)x^{-\lambda}e^{x}\delta(y-x)
+c_3\lim_{z\to1}\mathcal{J}(x,y,z)\right],
\end{eqnarray}
where 
\begin{eqnarray}
c_1=-\frac{\alpha\beta-\gamma\delta}{(\alpha+\gamma)^2(\beta+\delta)},
\quad c_2=\frac{\alpha}{\alpha+\gamma}, 
\quad c_3=-\frac{\alpha\beta-\gamma\delta}{(\alpha+\gamma)(\beta+\delta)},
\end{eqnarray}
\begin{eqnarray}
\mathcal{J}(x,y,z)&=&
\frac{1}{1-z}[xyz\mathcal{I}_{\lambda+1}(x,y,z)-yz\mathcal{I}_\lambda(x,y,z)],
\\
\mathcal{I}_\lambda(x,y,z)&=&
\sum_{n=0}^\infty \frac{n!}{\Gamma(n+\lambda+1)} 
L_n^{(\lambda)}(x) L_n^{(\lambda)}(y) z^n \nonumber\\
&=&\frac{(xyz)^{-\lambda/2}}{1-z}\exp \(-\frac{(x+y)z}{1-z}\)
I_\lambda\( \frac{2(xyz)^{1/2}}{1-z}\) .
\end{eqnarray}
Here $I_\lambda(z)$ is the modified Bessel function.

\setcounter{equation}{0}
\section{Partition Function $Z_L$ and Bulk Quantities}
\label{sec:Z_L}
Applying the formulae described in the previous sections, 
the partition functions of the ASEP, 
$Z_L$, are obtained in the integral form. The results are of course 
free from the choice of representation. 

\subsection{The case where $q\neq 1$}

We consider the representation in section \ref{sec:repAWq<1} since 
the boundary parameters are unrestricted. 
First, consider the $q<1$ case. 
Inserting the identity operator (\ref{eqn:orthop}) in  (\ref{eqn:defZ_L}), 
then we have 
\begin{eqnarray}
Z_L&=&
\llangle W\vert 
\C ^L
\vert V\rrangle 
\nonumber\\
&=&
h_0^{-1}
\oint_C\frac{dz}{4\pi iz} w\((z+z^{-1})/2\)
\llangle W\vert p\((z+z^{-1})/2\)\rangle 
\ \langle p\((z+z^{-1})/2\)\vert 
\C ^L
\vert V\rrangle 
\nonumber\\
&=&
h_0^{-1}
\oint_C\frac{dz}{4\pi iz} w\ 
\llangle W\vert p\((z+z^{-1})/2\)\rangle 
\ \langle p\((z+z^{-1})/2\) \vert V\rrangle 
\left[ \frac{(1+z)(1+z^{-1})}{1-q}\right] ^L 
\nonumber\\
&=&
\oint_C\frac{dz}{4\pi iz}
w\((z+z^{-1})/2\)
\left[ \frac{(1+z)(1+z^{-1})}{1-q}\right] ^L . \nonumber 
\end{eqnarray}
In the third line, we used (\ref{eqn:eigenp}). 
The integral formula of the partition function is now obtained: 
\begin{eqnarray}
Z_L=
\oint_C\frac{dz}{4\pi iz}
\frac{(z^2,z^{-2};q)_\infty
[(1+z)(1+z^{-1})/(1-q)]^L}
{(az,a/z,bz,b/z,cz,c/z,dz,d/z;q)_\infty} .
\label{eqn:Z_L}
\end{eqnarray}
where the integral contour $C$ is a closed path which 
encloses the poles at $z=aq^k$, $bq^k$, $cq^k$, $dq^k$$(k\in \Z_+)$ 
and excludes the poles at $z=(aq^k)^{-1}$, $(bq^k)^{-1}$, $(cq^k)^{-1}$, 
$(dq^k)^{-1}$$(k\in \Z_+)$. 
The partition function with the fugacity (\ref{eqn:Zfuga}) is similarly obtained 
with replacements 
$a\to a\xi^{-1},\ b\to b\xi ,\ c\to c\xi^{-1},\ d\to d\xi$: 
\begin{eqnarray}
Z_L(\xi^2)=
\oint_C\frac{dz}{4\pi iz}
\frac{(z^2,z^{-2};q)_\infty
[(1+\xi z)(1+\xi z^{-1})/(1-q)]^L}
{(a\xi^{-1}z,a\xi^{-1}/z,
b\xi z,b\xi /z,
c\xi^{-1}z,c\xi^{-1}/z,
d\xi z,d\xi/z;q)_\infty} .
\end{eqnarray}

The integral is hard to evaluate, but its asymptotic behavior 
in the thermodynamic limit $L\gg 1$ is obtained without difficulty. 
We should consider three cases: 
(A) $a>1$ and $a>b$, (B) $b>1$ and $b>a$ and (C) $a<1$ and $b<1$. 

\noindent 
(A) $a>1$ and $a>b$: In the integral, the dominant contribution comes 
from the residue of the poles at $z=a,a^{-1}$ of order $[(1+a)(1+a^{-1})/(1-q)]^L$. 
Therefore, we have an approximation of $Z_L$ as 
\begin{eqnarray}
Z_L^{(a)}\simeq 
\frac{(a^{-2};q)_\infty}
{(q,ab,ac,ad,a^{-1}b,a^{-1}c,a^{-1}d;q)_\infty}
\left[\frac{(1+a)(1+a^{-1})}{1-q}\right]^L .
\label{eqn:Z_L-A}
\end{eqnarray}
Similarly, for the partition function with fugacity,  we have 
\begin{eqnarray}
Z_L^{(a)}(\xi^2) \simeq 
\frac{(a^{-2}\xi^{2};q)_\infty}
{(q,ab,ac\xi^{-2},ad,a^{-1}b\xi^2,a^{-1}c,a^{-1}d\xi^2;q)_\infty}
\left[\frac{(1+a)(1+a^{-1}\xi^2)}{1-q}\right]^L.
\label{eqn:Z_L_xi-A}
\end{eqnarray}
Then, the bulk quantities 
(\ref{def:J}), (\ref{def:rho}) and (\ref{def:rho^2}) are 
calculated as 
\begin{eqnarray}
J\simeq 
(1-q)\frac{a}{(1+a)^2},\qquad
\langle \rho\rangle \simeq 
\frac{1}{1+a},\qquad
\langle \Delta\rho^2\rangle \simeq 
\frac{a}{(1+a)^2L}.
\end{eqnarray}

\noindent 
(B) $b>1$ and $b>a$: The results are similarly obtained as (A): 
\begin{eqnarray}
&&Z_L^{(b)}\simeq 
\frac{(b^{-2};q)_\infty}
{(q,ba,bc,bd,b^{-1}a,b^{-1}c,b^{-1}d;q)_\infty}
\left[\frac{(1+b)(1+b^{-1})}{1-q}\right]^L .
\label{Z_L-B}
\\
&&Z_L^{(b)}(\xi^2) \simeq 
\frac{(b^{-2}\xi^{-2};q)_\infty}
{(q,ba,bc,bd\xi^2,b^{-1}a\xi^{-2},b^{-1}c\xi^{-2},b^{-1}d;q)_\infty}
\left[ \frac{(1+b\xi^2)(1+b^{-1})}{1-q}\right]^L.
\end{eqnarray}
Then, we have 
\begin{eqnarray}
J\simeq (1-q)\frac{b}{(1+b)^2},\qquad
\langle \rho\rangle \simeq 
\frac{b}{1+b},\qquad
\langle \Delta\rho^2\rangle \simeq 
\frac{b}{(1+b)^2L}.
\end{eqnarray}

\noindent 
(C) $a<1$ and $b<1$: 
The contour integral is just over a unit circle 
and can be approximated by the steepest descent method: 
\begin{eqnarray}
Z_L^{(0)} \simeq 
\frac{(q;q)_\infty^2}{(a,b,c,d;q)_\infty^2}
\frac{4}{\sqrt{\pi}L^{3/2}}
\left(\frac{4}{1-q}\right)^L.
\label{Z_L-C}
\end{eqnarray}
\begin{eqnarray}
Z_L^{(0)}(\xi^2) \simeq 
\frac{(q;q)_\infty^2}{(a\xi^{-1},b\xi ,c\xi^{-1},d\xi ;q)_\infty^2 }
\frac{[(1+\xi)(1+\xi^{-1})]^{3/2}}{2\sqrt{\pi}L^{3/2}}
\left[ \frac{(1+\xi)^2}{1-q}\right] ^L . 
\end{eqnarray}
Then, we have 
\begin{eqnarray}
J\simeq 
\frac{1-q}{4},\qquad
\langle \rho\rangle \simeq 
\frac{1}{2},\qquad
\langle \Delta\rho^2\rangle \simeq 
\frac{1}{8L}.
\end{eqnarray}
The results for $q>1$ are directly from the above expressions 
considering the reflection symmetry (\ref{eqn:rs1}). 

\noindent
(A') $b^{-1}>1$ and $b^{-1}>a^{-1}$:
\begin{eqnarray}
J\simeq 
-(1-q^{-1})\frac{b}{(1+b)^2},\qquad
\langle \rho\rangle \simeq 
\frac{b}{1+b},\qquad
\langle \Delta\rho^2\rangle \simeq 
\frac{b}{(1+b)^2 L}.
\end{eqnarray}

\noindent 
(B') $a^{-1}>1$ and $a^{-1}>b^{-1}$: 
\begin{eqnarray}
J\simeq -(1-q^{-1})\frac{a}{(1+a)^2},
\qquad
\langle \rho\rangle \simeq 
\frac{1}{1+a},\qquad
\langle \Delta\rho^2\rangle \simeq 
\frac{a}{(1+a)^2 L}.
\end{eqnarray}

\noindent 
(C') $a^{-1}<1$ and $b^{-1}<1$: 
\begin{eqnarray}
J\simeq 
-\frac{1-q^{-1}}{4},\qquad
\langle \rho\rangle \simeq 
\frac{1}{2},\qquad
\langle \Delta\rho^2\rangle \simeq 
\frac{1}{8L}.
\end{eqnarray}
The minus sign in the front of the current is 
due to the parity change under the reflection. 

\begin{figure}[htb]
\begin{center}
\setlength{\unitlength}{1mm}
\begin{picture}(150,150)
\put(20,20){\vector(1,0){105}}
\put(20,20){\vector(0,1){105}}
\put(20.5,20){\line(1,1){50}}
\put(20,20.5){\line(1,1){50}}
\put(70,70){\line(1,0){50}}
\put(70,70){\line(0,1){50}}
\put(40,90){{\Huge \bf{A}}}
\put(90,40){{\Huge \bf{B}}}
\put(90,90){{\Huge \bf{C}}}
\put(120,10){{\Huge $x$}}
\put(10,120){{\Huge $y$}}
\put(70,10){{\Huge $1$}}
\put(5,70){{\Huge $1$}}
\end{picture}
\end{center}
\caption{The phase diagram of the ASEP. 
There are three phases; 
phase A: low-density phase, phase B: high-density phase 
and phase C: maximal-current phase. 
$(x,y)=(a^{-1},b^{-1})$ if $q<1$, and $(x,y)=(b,a)$ if $q>1$. 
$a$ and $b$ are defined in (\ref{abcd}). 
The double line is for the transition of the first order 
and the solid line is for that of the second order.}
\label{fig:PhaseD}
\end{figure}
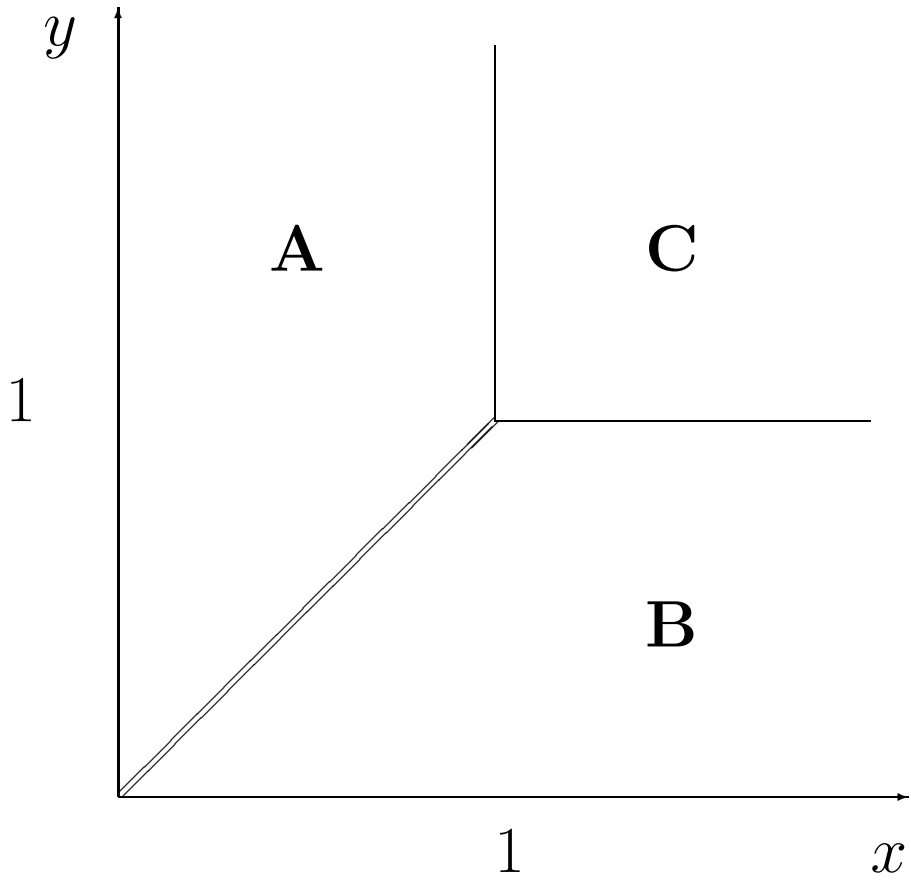

In this way, we have three phases altogether. 
The phase A is called the low-density phase, 
the phase B the high-density phase, 
and the phase C the maximal-current phase. 
In fact, the average of particle density increases in the order of 
phase A, phase C, phase B. 
The particle current takes the maximum value in phase C. 
Figure \ref{fig:PhaseD} shows the phase diagram of $J$ and $\langle\rho\rangle$. 
The same diagram as that in \cite{Sandow94} is reproduced. 
We observe that across the line between the phase A and B, 
the first derivative of $J$ is discontinuous 
and $\langle\rho\rangle$ itself is discontinuous. 
Across the lines between phase A and C, or phase B and C, 
the second derivative of $J$ is discontinuous 
and the first derivative of $\langle\rho\rangle$ is discontinuous. 
Notice that $J$ is given by the partition function $Z_L$ 
and $\langle\rho\rangle$ is given by the first derivative of $Z_L$. 
Then, in the terminology of statistical mechanics, we can say 
the transition is of first order across the line between phase A and B, 
and is of second order across the line between phase A and C, or phase B and C. 
It is interesting to discuss some extreme limit cases. 
When $q>1$ and $\delta=0$, or equivalently $d=0$, the situation is that 
there is no particle supply at the right boundary. 
Since the particles are likely to move leftward in the bulk 
and can move out the system at the left boundary, 
the system becomes short of particles in the end. 
In this extreme case, as remarked in the end of section \ref{sec:AW} 
the weight function in the integral should be changed into the form of (\ref{eqn:-1w}). 
The partition function $Z_L$ is then 
\begin{eqnarray}
Z_L&=&\int_{-\infty}^\infty
\frac{du}{\log q}
\frac{(iaq^{-1}e^u,-iaq^{-1}e^{-u},
ibq^{-1}e^u,-ibq^{-1}e^{-u},
icq^{-1}e^u,-icq^{-1}e^{-u};\Q)_\infty}
{(-q^{-1}e^{2u},-q^{-1}e^{-2u};q^{-1})_\infty}
\nonumber\\
&&\qquad\times\left[ \frac{2+2i\sinh u}{1-q} \right]^L. 
\end{eqnarray}
Similarly, when $q>1$ and $\gamma=0$, or equivalently $c=0$, the situation is that 
there is no escape of particles at the left boundary. 
More and more particles accumulate near the left "blocking" boundary and 
finally the system is stopped with jammed particles. 
The partition function in this case is almost the same as the above. 
Just replace $c$ by $d$. 
We remark that the case where $q>1$ and $c=d=0$ (or, $\gamma=\delta=0$)
was treated in \cite{BECE}.

\subsection{The case where $q=1$}

The partition function for $q=1$ is given by
\begin{eqnarray}
Z_L&=&\frac{1}{\Gamma(\lambda+1)}
\int_0^\infty \!\! dx\ \ x^{\lambda+L}e^{-x} \nonumber\\
&=&\frac{\Gamma(\lambda+L+1)}{\Gamma(\lambda+1)}.
\end{eqnarray}
The current is calculated as
\begin{eqnarray}
J=\frac{1}{\lambda+L}.
\end{eqnarray}
Instead of the average density (\ref{def:rho}), 
we compute exactly the particle density profile 
$\langle \tau_k \rangle$ in the next section.

\setcounter{equation}{0}
\section{The $n$-Point Function}
\label{sec:nPointFunction}

The $n$-point functions are obtained in the integral form as well. In general, 
we find that the $n$-point function contains $n+1$-multiple integrals.

\subsection{The case where $q\neq 1$}

Inserting the identity operator (\ref{eqn:orthop}) $n+1$ times altogether 
between $\D$'s in the matrix products expression
and using the form factor (\ref{eqn:formfactor}), 
we obtain the $n$-point function for $j_1<\cdots<j_n$, 
\begin{eqnarray}
&& \langle \tau_{j_1}\cdots\tau_{j_n}\rangle=
\frac{1}{(1-q)^L Z_L}
\lim_{\xi_1,\cdots,\xi_n\to1} 
\left[ \prod_{m=1}^{n+1} \oint_{C_m} \frac{dz_m}{4\pi iz_m} \right] \nonumber\\
&&\qquad\times
	\frac{\displaystyle{\prod_{m=1}^{n+1}(z_m^2,z_m^{-2};q)_\infty
	[(1+z_m)(1+1/z_m)]^{j_m-j_{m-1}-1}}}
	{(az_1,a/z_1,cz_1,c/z_1,
	bz_{n+1},b/z_{n+1},dz_{n+1},d/z_{n+1};q)_\infty} 
\nonumber\\
&&\qquad\times
	\prod_{m=1}^{n}
	\bigg[
	\frac{1}{(z_m^2,z_m^{-2};q)_\infty} \delta(z_{m+1}-z_m)
\nonumber\\
&&\qquad\qquad+
	\frac{(q;q)_\infty^2 (z_{m+1}+1/z_{m+1}-\xi_m z_m-\xi_m /z_m)}
	{(\xi_m z_mz_{m+1},\xi_m z_m/z_{m+1},
	\xi_m z_{m+1}/z_m,\xi_m /z_mz_{m+1};q)_\infty}\bigg]. 
\end{eqnarray}
where $j_0=0,j_{n+1}=L+1$ in the product. 
Regularization parameters $\xi_m$'s are included since otherwise this integral 
becomes a singular integral. 
The contours $C_m(m=2,\cdots,n+1)$ encloses the poles at 
$z_m=\xi_{m-1} z_{m-1}q^k,$ $\xi_{m-1} z_{m-1}^{-1}q^k$ $(k\in\Z_+)$ 
and excludes the poles at 
$z_m=(\xi_{m-1} z_{m-1}q^k)^{-1},$ $(\xi_{m-1} z_{m-1}^{-1}q^k)^{-1}$ $(k\in\Z_+)$. 
In addition, for $m=1$, $C_1$ encloses the poles at 
$z_1=aq^k,cq^k$ $(k\in\Z_+)$ and excludes the poles at 
$z_1=(aq^k)^{-1},(cq^k)^{-1}$ $(k\in\Z_+)$. 
Also, $C_{n+1}$ encloses the poles at $z_{n+1}=bq^k,dq^k$ $(k\in\Z_+)$ 
and excludes the poles at $z_{n+1}=(bq^k)^{-1},(dq^k)^{-1}$ $(k\in\Z_+)$. 
Note that the result should be independent of the order of 
inserting of the identity operator. 
The above choice of integral contours is when the insertions are done 
from left to right.

The results for the $q>1$ case are simply obtained from the 
reflection symmetry (\ref{eqn:rs1}).

\subsection{The case where $q=1$}

Applying formulae in section \ref{sec:repq=1}, 
we also obtain the $n$-point function for $q=1$: 
\begin{eqnarray}
&& \langle \tau_{j_1}\cdots\tau_{j_n}\rangle=
	\frac{1}{\mu Z_L}
	\lim_{\xi_1,\cdots,\xi_n\to1} 
	\left[ \prod_{m=1}^{n+1} \int_0^\infty \!\!\!
	dx_m \ x_m^\lambda e^{-x_m} \right] 
	\prod_{m=1}^{n+1} x_m^{j_m-j_{m-1}-1} 
\nonumber\\
&&\qquad\times
	\prod_{m=1}^{n}
	\left[(c_1+c_2x_m)x_m^{-\lambda}e^{x_m}\delta(x_{m+1}-x_m)
	+c_3\mathcal{J}(x_m,x_{m+1},\xi_m)\right],
\end{eqnarray}
where $j_0=0,j_{n+1}=L+1$ in the product. 
In particular, for the one-point function the integral can be performed directly. 
Thus, we obtain 
\begin{eqnarray}
\langle \tau_k \rangle =
\frac{\alpha}{\alpha+\gamma}-
\frac{1}{\lambda+L}\frac{\alpha\beta-\gamma\delta}{(\alpha+\gamma)(\beta+\delta)}
\(\frac{1}{\alpha+\gamma}+k-1\) ,
\label{eqn:tau_k}
\end{eqnarray}
for $k=1,\cdots,L$. 
This shows a linear profile of particle density.

\section{Conclusion}
\label{sec:conclusion}

We have studied the steady state of the ASEP for the most general 
case of open boundary conditions where particles hop in both directions 
and are injected and ejected at both boundaries.
Finding out an explicit representation of the algebraic relations 
related to the Askey-Wilson polynomials, we have obtained the 
partition function in the form of the moments with respect to 
the weight function of the Askey-Wilson polynomials. 
This is an extension of the previous work for two degrees 
of freedom to the most general case. 
Furthermore, we also have found 
a novel integral formula for the $n$-point functions. 

From the asymptotic behaviors of the partition function,
the current in the thermodynamic limit has been calculated.
The phase diagram is given and the
boundary-induced phase transitions are observed.
For the $n$-point functions, a limit procedure is included in the 
multiple integrals. 
The asymptotic evaluation of them is more difficult, 
but could be done by generalizing the methods in \cite{SS2000}.

The advantage of our method using the orthogonal polynomials 
is that, once the corresponding polynomials are known, 
one can use the whole storage of knowledge on them. 
It should be stressed that the Askey-Wilson polynomials are 
considered to be one of the most important orthogonal polynomials 
because all important classical orthogonal polynomials are 
obtained as special/limiting cases of the Askey-Wilson polynomials.

The discovery of the connection to the Askey-Wilson 
polynomials is yet another evidence of a deep 
mathematical structure of the ASEP, which has been already
suggested in several previous works.
For instance, the ASEP can be considered as an exactly 
solvable spin system and  the Bethe ansatz methods has been applied 
successfully \cite{Schu93,Schu97}. 
The connection of the matrix method to the 
Zamolodchikov-Faddeev algebra was pointed out in \cite{SSW97}.
In addition, more recently, fluctuations of the current of 
the ASEP have been studied using the connection to the random matrix
theory \cite{Jo2000,PS2002a}.
It would be of great interest to further clarify the interrelationship
among these fascinating connections.

\pagebreak

\pagebreak

%%%%% caps of figs 

\begin{figure}[htb]
\begin{center}
\setlength{\unitlength}{1mm}
\begin{picture}(100,15)
\multiput(10,0)(10,0){8}{\framebox(10,10)}
\put(7,7){\vector(1,0){6}}
\put(13,3){\vector(-1,0){6}}
\put(87,7){\vector(1,0){6}}
\put(93,3){\vector(-1,0){6}}
\put(35,5){\circle*{4}}
\put(55,5){\circle*{4}}
\put(65,5){\circle*{4}}
\put(15,5){\circle*{4}}
\put(75,5){\circle*{4}}
\put(32,5){\vector(-1,0){5}}
\put(38,5){\vector(1,0){5}}
\put(3,7){$\alpha$}
\put(3,2){$\gamma$}
\put(95,7){$\beta$}
\put(95,1){$\delta$}
\put(25,7){$p_L$}
\put(43,7){$p_R$}
\put(11,-5){$i=1$}
\put(25,-5){$2$}
\put(52,-5){$\cdots$}
\put(84,-5){$L$}
\end{picture}
\end{center}
\caption{The ASEP with open boundaries}
\label{asepfig}
\end{figure}
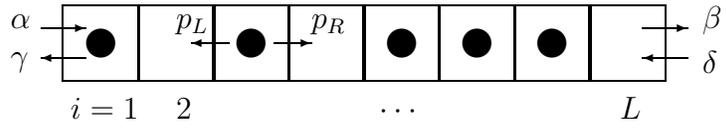

\end{document}